\documentclass[12pt]{article}

\usepackage{newtxtext,newtxmath}

\usepackage{graphicx}

\usepackage[letterpaper,margin=1in]{geometry}

\renewcommand{\implies}{\Rightarrow}
\newcommand{\entails}{\vdash}
\newcommand{\algebraicset}[1]{{\kappa(#1)}}
\newcommand{\sls}[1]{\mathcal{L}_{#1}}
\newcommand{\receiverrndm}{ \mathtt{R_m} }
\newcommand{\sender}{ \mathtt{s} }
\newcommand{\senderrndhat}{ \mathtt{\hat{S}_m}}
\newcommand{\senderrndm}{ \mathtt{S_m}}
\newcommand{\queryrndm}{\mathtt{Q_m}}
\newcommand{\queryrndmapo}{\mathtt{Q_m^{'}}}

\newtheorem{theorem}{Theorem}

\renewenvironment{abstract}
	{\quotation}
	{\endquotation}

\date{}

\makeatletter
\renewcommand{\fnum@figure}{\textbf{Figure \thefigure}}
\renewcommand{\fnum@table}{\textbf{Table \thetable}}
\makeatother

\usepackage{scicite}

\usepackage{url}

\def\scititle{
Breaking through the classical Shannon entropy limit: A new frontier through logical semantics
}
\title{\bfseries \boldmath \scititle}

\author{
	Luis~A.~Lastras$^{1\ast}$,
	Barry~M.~Trager$^{1}$,
	Jonathan~Lenchner$^{1}$\and
    Wojciech~Szpankowski$^{2}$, 
    Chai~Wah~Wu$^{1}$, 
    Mark~S.~Squillante$^{1}$, 
    and Alexander~Gray$^{3,2}$ \and
	\small$^{1}$IBM Research AI, Yorktown Heights, NY, 10598, USA.\and
	\small$^{2}$Purdue University, West Lafayette, IN, 47907, USA.\and
    \small$^{3}$Centaur AI Institute\and
	\small$^\ast$Corresponding author. Email: lastrasl@us.ibm.com
}

\begin{document} 

\maketitle

\begin{abstract} \bfseries \boldmath
Information theory has provided  foundations for the theories of several application areas critical for modern society, including communications, computer storage, and AI.  A key aspect of Shannon's 1948 theory is a sharp lower bound on the number of bits needed to encode and communicate a string of symbols. When he introduced the theory, Shannon famously excluded any notion of semantics behind the symbols being communicated. This semantics-free notion went on to have massive impact on communication and computing technologies, even as multiple proposals for reintroducing semantics in a theory of information were being made, notably one where Carnap and Bar-Hillel used logic and reasoning to capture semantics. In this paper we present, for the first time, a Shannon-style analysis of a communication system equipped with a deductive reasoning capability, implemented using logical inference. We use some of the most important techniques developed in information theory to demonstrate significant and sometimes surprising gains in communication efficiency availed to us through such capability, demonstrated also through practical codes. We thus argue that proposals for a semantic information theory should include the power of deductive reasoning to magnify the value of transmitted bits as we strive to fully unlock the inherent potential of semantics.
\end{abstract}

\noindent
At the beginning of his Lectures on Physics \cite{feynman_1965_flp}, Feynman posed a hypothetical situation where all scientific knowledge is destroyed, and a statement that has the most information in the fewest words needs to be chosen. His choice was ``all things are made of atoms'', which assumes that scientists would be able, through experimentation, induction and deduction, to reconstruct vast amounts of our scientific knowledge from the one statement.

Feynman's notion of information agrees with most accepted definitions of the word. Yet, in 1948, when Shannon founded the modern field of information theory,  he famously argued that only the statistical patterns of the symbols being sent were relevant to the engineering problem of communication, and not any notion of the meaning, or {\it semantics}, of the symbols \cite{shannon:BST48}. In the absence of a deductive process, effectively conveying all scientific knowledge would seem to require a large quantity of bits to be transmitted. Yet Feynman's sentence can be transmitted with just a few dozen bits.

Shannon, of course, knew exactly what he was doing, as he had a very specific goal in mind. By all accounts, Shannon's perspective has prevailed for decades, as it was immediately consequential, having helped launch a revolution in digital communications that has given us extraordinary means for exchanging information efficiently without having to worry about the intent or meaning behind our messages during the design of such systems.

{\bf A new frontier.} It may be, nonetheless, that the time has come to deepen the reach of information theory \cite{ayg-spa} towards the realm of semantics. This frontier has always been there.  In 1952, Carnap and Bar-Hillel \cite{carnap53} utilized the long-standing formalization of semantics as {\em logic}, which can be traced back from Aristotle (logic as syllogism) through to Boole (Boolean logic) and Frege (First-Order Logic), to propose a mathematical conceptualization of a semantic information theory. Shannon himself \cite{shannon:lattice} argued that entropy ``can hardly be said to represent the actual information'' in a little-known paper introducing lattices as part of a more general theory of information. 
In both \cite{carnap53, shannon:lattice} we find the idea that if a statement can be deduced from the  statements one already knows, it conveys zero information, but the more general implications  of this observation to communication systems were left unexplored. A number of proposals have been made in the interim \cite{floridi:semantic, Devlin1991-DEVLAI, basu:semantic, basu:semantic_extended, liu:rd_semantic, liu_shao_zhang_poor:indirect_rd_semantic, Shao2022ATO, guo2022semantic, stavrou:goal_semantic_rd,gunduz:beyond_bits, niu2024mathematicaltheorysemanticcommunication, yu:information_lattice_learning, Yu2024SemanticCW}, which roughly focus on either the fundamental aspects of defining a notion of semantic information, matters relating to transmission of semantic information, or problems of learning models of semantic information. 
In a subset of works that have deeper connections to classical information theory \cite{liu:rd_semantic,liu_shao_zhang_poor:indirect_rd_semantic, Shao2022ATO, guo2022semantic, stavrou:goal_semantic_rd, gunduz:beyond_bits,niu2024mathematicaltheorysemanticcommunication}, a common thread is the exploitation of Rate-Distortion theory \cite{shannon:rd}, which Shannon introduced to extend the lossless compression results of \cite{shannon:BST48} to lossy coding \cite{berger1971rate}. In these works, semantics are introduced by asserting that the fidelity criterion in lossy coding is of a semantic nature; they have not, nonetheless, explicitly considered reasoning in their frameworks. A different line of investigation is offered by \cite{yu:information_lattice_learning,Yu2024SemanticCW} who start from Shannon's information lattice paper \cite{shannon:lattice}. It is fair to say that the field of semantic information theory has yet to have the kind of impact that its classic counterpart has enjoyed to-date. We thus state that this frontier remains new and ripe to be explored.

{\bf The power of deduction. }
Logical deduction inherently has the powerful combinatorial capability to represent massive sets compactly, through use of a deductive reasoning engine which can chain together principles and facts. The assumption that a receiver has a deductive reasoning ability  should allow for more compact messages to be sent between the parties. Our purpose is to put this intuition under a rigorous mathematical magnifying glass. 

{\bf Our contribution.} Building on top of the foundations established by Carnap and Bar-Hillel \cite{carnap53}, model theoretic logics \cite{barwise-model-theoretic-logics-1985}, Shannon's Rate-Distortion theory \cite{shannon:rd},  and the theories of source coding with side information due to Slepian-Wolf \cite{slepian_wolf:coding} and Wyner-Ziv \cite{wyner_ziv:coding} coding, we provide, for the first time, a rigorous theory that incorporates deductive reasoning directly in the communication process, providing sharp upper and lower bounds on communication cost under a wide variety of scenarios often showing significant efficiency gains compared to classic approaches. We also provide preliminary evidence of practical systems realizing a fraction of these possible gains. Our results hold for deductive mechanisms based on logic that is strongly sound and exhibits a type of strong completeness; we illustrate here using  propositional logic for didactic reasons. Insights are offered in three key settings, where: the sender is unaware of what the receive knows; there is partial information sharing; or there is misinformation.

\subsection*{Mathematical setup}

\begin{figure}
	\centering
\includegraphics[width=\textwidth]{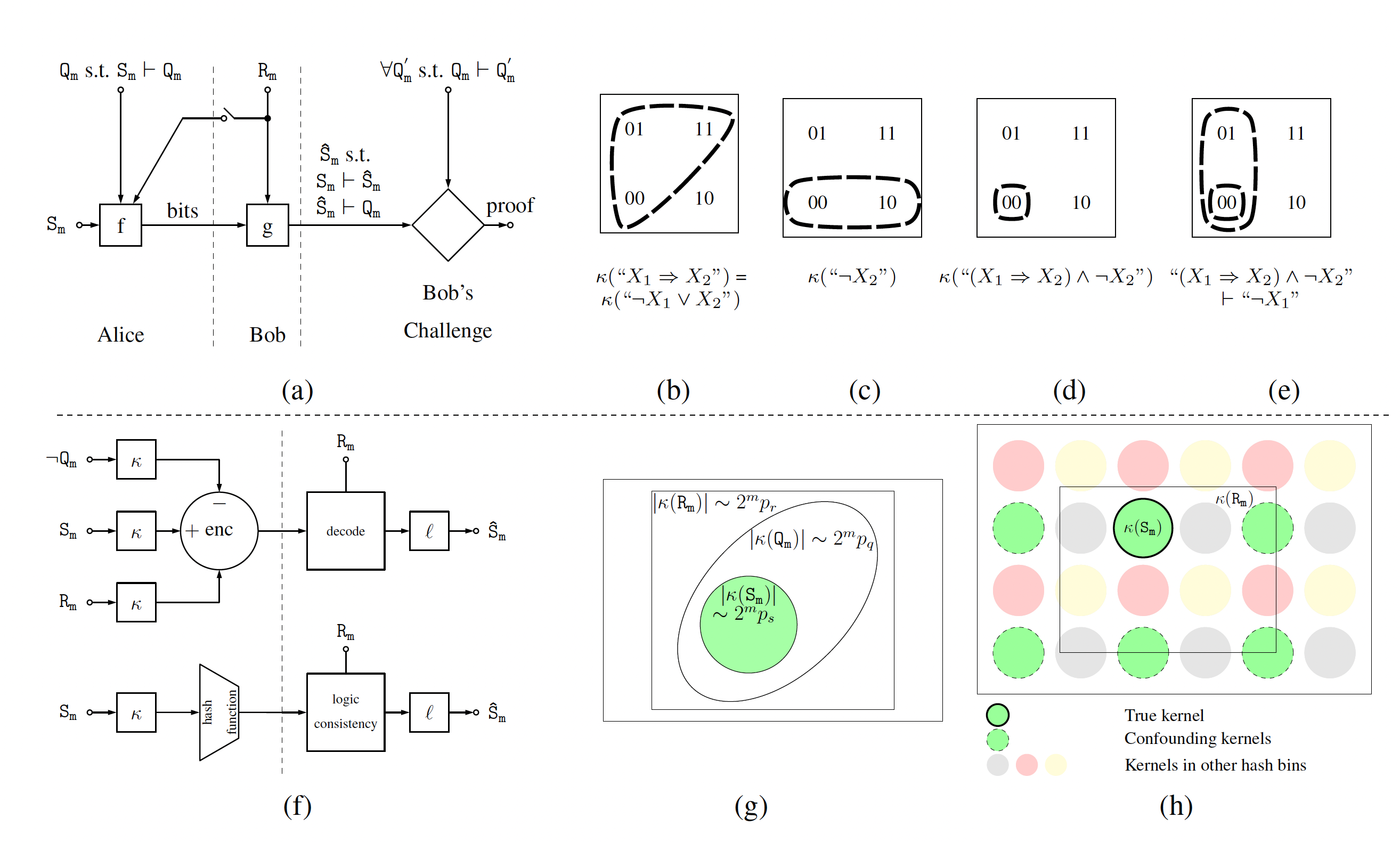}
\caption{\textbf{Communication problem setup and algorithm innovations.} (\textbf{a}) \emph{General communication diagram} which allows for Bob to possess facts that Alice may or may not know, and which allows for the query $\queryrndm$ that Alice is helping Bob prove to be anything from her full knowledge $\senderrndm$ to a more targeted logic statement. 
(\textbf{b})-(\textbf{e}) \emph{Kernel diagrams} showing equivalence, not, conjunction, and entailment. 
(\textbf{f}) \emph{Theoretically optimum solution architectures} for the cases where Alice knows and does not know Bob's knowledge, respectively.
(\textbf{g}) \emph{Probabilistic model for kernels} where the parameters $p_s, p_q, p_r$ are average kernel sizes, illustrated in the case Alice's knowledge implies that of Bob's. 
(\textbf{h}) \emph{No need to know} decoding mechanism example where Alice mapped her kernel  to a hash bin from 4 possible ones; Bob is able to reconstruct $\algebraicset{\senderrndm}$ without Alice knowing $\receiverrndm$ ($p_r < 1$) by rejecting all kernels that do not entail $\receiverrndm$ and that do not match the received bin index.}
\label{fig:e2e}
\end{figure}

Our first task at hand is to introduce in a classical communication system a computational device that implements a deductive reasoning process. We refer the reader to Figure~\ref{fig:e2e}(a), where the sender is Alice and the receiver is Bob; the reasoning device is illustrated as a rhombus. Given a starting statement (``cows eat grass and my pet does not eat grass'') and a query to be proved (``my pet is not a cow''), this device is able to either prove the query or conclude that the query cannot be proved from the starting logic statement.

Our second task is to specify a logic system. Throughout this article, for didactic purposes, we exemplify our work in the context of propositional logic with $m$ ordered propositional variables $\mathtt{X_1,\ldots,X_m}$ with binary assignments; logic statements about these propositions are strings from a logical language $\sls{m}$, using the symbols $\{ \land, \lor, \lnot, \implies, (, )\}$, in addition to the variables. Our work nonetheless is rigorously valid for many other logics; see~\cite{OurArxiv} for details. We use the symbol $\entails$ to denote when one logic statement entails another logic statement, i.e., the statement on the right of the symbol can be inferred from the statement on its left. 
We define the kernel $\algebraicset{\sender}$ of a logic statement $\sender \in \sls{m}$ as the set of values for the propositional variables that make such a statement true. Figures~\ref{fig:e2e}(b)-(d) depict three such kernels in the case of $m=2$. Figure~\ref{fig:e2e}(b) also illustrates how two different statements that are logically equivalent have the same kernel, and Figure~\ref{fig:e2e}(d) shows how the kernel of a conjunction of statements is the intersection of the individual kernels. Figure~\ref{fig:e2e}(e)  illustrates an idea of central importance to our article: a logic statement entails another logic statement if and only if the kernel of the former is a subset of that of the latter. While these are shown as illustrations, they are all general mathematical facts. The concept of a kernel in the context of a semantic information theory traces its roots back to Carnap and Bar-Hillel, who defined a closely related concept using the word ``range''.

{\bf Initial meeting.}
Alice and Bob meet ahead of time, and agree that the goal is for Bob to prove the truth of a logic statement $\queryrndm \in \sls{m}$ using information that Alice will provide employing a pre-agreed upon encoding. The query $\queryrndm$ is not known at the time of this meeting, and will be revealed to Alice later. The scenario where the query is revealed to Bob instead is possible and left to future work.

{\bf Correlated world observations.}
After this meeting, Alice and Bob go their own ways; Alice, the sender, obtains knowledge about the world summarized in a logic statement $\senderrndm \in \sls{m}$ whereas Bob, the receiver, obtains $\receiverrndm \in \sls{m}$. We consider both settings where Alice knows and doesn't know $\receiverrndm$; we do assume that $\senderrndm \entails \receiverrndm$ except in an interesting misinformation scenario discussed later.  This entailment assumption roughly corresponds to a situation where Alice has made a superset of the observations about the world that Bob has made (but may not know exactly what subset Bob has), and where each have independently encoded their observations as logic statements. The query $\queryrndm$ is also revealed to Alice at this time; we assume that $\senderrndm \entails \queryrndm$; in words, Alice will help Bob prove something she can prove with her knowledge.  In the case Alice does not know $\receiverrndm$, we only present the case where Alice is equipping Bob to prove all she can ($\queryrndm=\senderrndm$). In the case Alice does know $\receiverrndm$, we assume that $\queryrndm \entails \receiverrndm$. This last assumption is made without any loss of essential generality due to the fact that both Alice and Bob know $\receiverrndm$. For the full argument, we refer the reader to \cite{OurArxiv}. All three of $\senderrndm, \queryrndm, \receiverrndm$ are regarded as random; the assumptions on their distribution will be described shortly.

{\bf Communication.}
 In the case Alice knows $\receiverrndm$, she uses an encoder $f(\queryrndm, \senderrndm, \receiverrndm)$ to generate  bits that are transmitted to Bob. In turn, Bob uses a decoder $g(\receiverrndm,f(\queryrndm, \senderrndm, \receiverrndm))$ to obtain $\senderrndhat$, which represents Bob's updated logic statement. In all cases we assume that Bob ends up with knowledge consistent with that of Alice's, written mathematically as $\senderrndm \entails \senderrndhat$ (but the reverse entailment may not always hold). In the more challenging setting where Alice doesn't know $\receiverrndm$ (and we restrict $\queryrndm=\senderrndm$), we allow for multiple communication rounds where Alice and Bob exchange turns in sending information, culminating again with Bob's updated logic statement~$\senderrndhat$. In either case, we use exactly the same figure of merit: the total number of expected bits spent during the entire communication, no matter which direction, where the expectation is with respect to $\senderrndm, \queryrndm, \receiverrndm$.
This setup has some similarities with Yao's communication complexity \cite{10.1145/800135.804414}, one  difference is that we do not identify ahead of time a function both Alice and Bob want to compute.

 {\bf Challenge and deduction.}
After the communication takes place, Bob is challenged with any statement $\queryrndmapo$ that can be proven by $\queryrndm$ (including possibly $\queryrndm$ itself), and Bob is able to produce a proof for that query using the logical sentences  $\senderrndhat$ and $\receiverrndm$; mathematically, for the system to have succeeded, it must be the case that $\senderrndhat \entails \queryrndm$.

{\bf Probabilistic model.}
We will present a theoretical result in the form of upper and lower bounds on the total number of expected bits. Our upper bounds are applicable to any possible distribution over $\senderrndm, \queryrndm, \receiverrndm$ as long as the entailment conditions described earlier are met; these are illustrated geometrically in Figure~\ref{fig:e2e}(g). The results are phrased in terms of normalized versions of the expected kernel sizes.  As we are assuming binary valued propositions, the total number of possible values that $m$ propositions can take is $2^m$. The normalized sizes of the kernels in Figure~\ref{fig:e2e}(b),(c),(d) are 3/4, 1/2 and 1/4, respectively. Given any probability distribution, we denote the expected normalized sizes of the kernels of $\senderrndm, \queryrndm$ and $\receiverrndm$ as $p_s, p_q$ and $p_r$, respectively, where $p_s \leq p_q < p_r \leq 1$. 
For our lower bounds, we use a stronger assumption  that can be roughly described as one where the elements of a kernel are chosen 
independent and identically distributed
(i.i.d.)
with some probability, from the elements of a subsuming kernel, resulting in the same overall expected sizes as before; in this case, these three numbers can be more readily interpreted as probabilities. The precise definitions can be found in~\cite{OurArxiv}.
In the case Alice does not know Bob's $\receiverrndm$, the model by assumption is simpler:  only $\senderrndm$ and $\receiverrndm$ are relevant; we thus set $\queryrndm=\senderrndm$ and $p_q=p_s$. 

\subsection*{Overview of results}
All of our results are phrased in terms of a type of scaled conditional entropy that we call the {\it logical semantic entropy}, denoted by $\Lambda$, which increases monotonically in each variable:
\begin{eqnarray}
    \Lambda(a,b) = a \log_2 \left( \frac{a+b}{a} \right) + b \log_2 \left( \frac{a+b}{b} \right) .
\end{eqnarray}
\noindent The discovery of the role of this function in logical semantic communication is a key part of our contribution. Our main result, to be interpreted in the context of the above setup, is stated next.
\begin{theorem}
For any distribution over $(\senderrndm, \queryrndm, \receiverrndm)$ meeting the entailment conditions $\senderrndm \entails \queryrndm$ and $\queryrndm \entails \receiverrndm$, if the corresponding kernels have normalized sizes $p_s,p_q,p_r$, respectively, then for the case Alice knows $\receiverrndm$, an algorithm exists with normalized average cost in total bits exchanged that is upper bounded by $\Lambda(p_s,p_r-p_q) + O(m / 2^m)$. Under an additional ``i.i.d.'' constraint (see \cite{OurArxiv}), the normalized average cost of any such algorithm is lower bounded by $\Lambda(p_s,p_r-p_q)$. In the case Alice does not know $\receiverrndm$, under the additional assumption that $\queryrndm=\senderrndm$ and thus $p_q=p_s$, the same conclusions hold.
\label{thm:core}
\end{theorem}

This results holds more generally beyond Propositional Logic - see \cite{OurArxiv}.

    {\bf Solution architecture.} In Figure~\ref{fig:e2e}(f), we illustrate two architectures that achieve the upper bound in this theorem in the cases where Alice knows and doesn't know $\receiverrndm$. In both cases, we map the sender's logic statement to its kernel. When Alice does know $\receiverrndm$ and the query may be more targeted, the key technique involves approximating that kernel with another which can also prove $\queryrndm$, chosen from a codebook of limited size, illustrated as the circle labeled `enc'. When Alice does not know $\receiverrndm$, the key idea is compressing Alice's kernel by sending only the index of a bin (hash bin) to which it is mapped by hashing. 
   Bob is able to recover the approximating kernel by choosing from the hash bin the one kernel that entails $\receiverrndm$, thus eliminating confounding ones (Figure~\ref{fig:e2e}(h)). To ensure this succeeds with very high probability while keeping very good compression, the hash bins need to be sized ``just right''. This is attained by multiple preliminary communication rounds where sender and receiver communicate their kernel sizes; the vanishingly small probability event where the protocol runs into difficulties is addressed with one optional final round where the sender kernel is transmitted with a simple encoding. The process ends by using a function $\ell$ that is able to translate a kernel back to an exemplary logic statement $\senderrndhat$. Arguing why and how these techniques result in such an optimal systems is beyond the scope of this paper and fully addressed in \cite{OurArxiv}.

\begin{figure}
	\centering
\includegraphics[width=\textwidth]{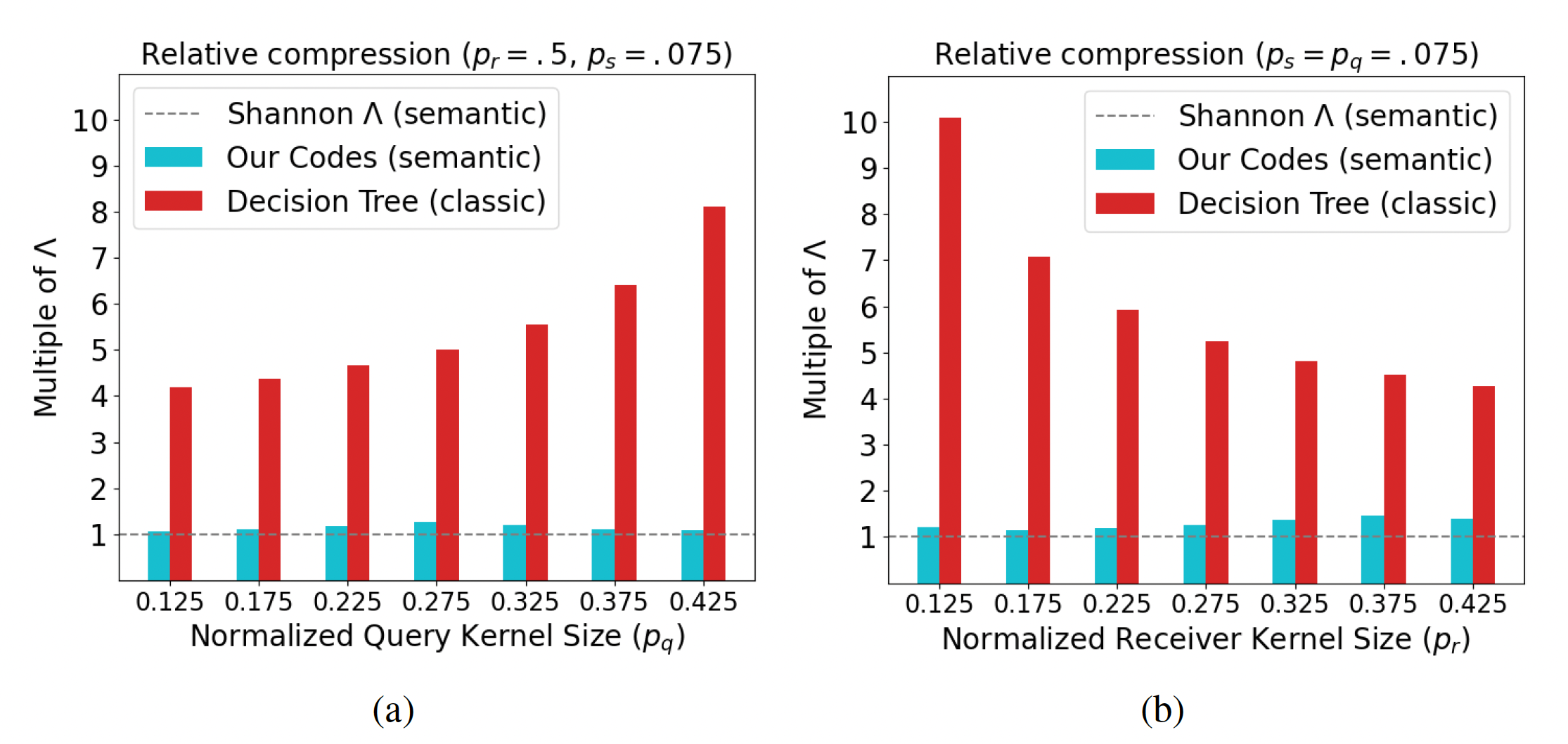} 
\caption{ \textbf{Experimental results for two scenarios.}
(a) Results for the case Alice knows $\receiverrndm$, $p_r=0.5, p_s=0.075$ and $0.125 \leq p_q \leq 0.425$ demonstrating the significant gains in communication cost using practical semantic communication codes compared to purely classical approaches. (b) Results for the setting $p_r < 0.5$ and $p_s=p_q=0.075$ in the case Alice doesn't know $\receiverrndm$, also comparing a classical approach with one leveraging logical semantics, as a multiple of the Shannon bound.}
\label{fig:empirical_results}
\end{figure}

{\bf Empirical validation.} In Figure~\ref{fig:empirical_results}, we present results of practical codes for two contrasting situations. On the left, Alice and Bob have access to $\receiverrndm$, and Alice will help Bob prove a statement $\queryrndm$ narrower than $\senderrndm$,  parametrized by the average normalized kernel size $p_q$. The red bars  correspond to a coding technique based on an optimized representation  (``decision trees'', explained 
in~\cite{OurArxiv}) 
of either the sender or query logic statements, whichever is cheapest after being passed through a good off-the-shelf lossless compressor (classic compression). The light blue bars correspond to practical codes for semantic logic communication that we have developed as part of this work, fully described in~\cite{OurArxiv}.
Both are plotted relative to the ultimate information-theoretic bound $\Lambda$, shown by the dashed line. Our practical codes use a combination of a novel application of linear codes with the exploitation of classical techniques such as Cover's enumerative source coding \cite{cover:enumerative} and integer coding techniques such as Elias' \cite{elias:integers}. 

To implement the competing classic approaches, we developed an optimized representation of logic statements based on decision trees \cite{breitbart:size_of_binary_decision_diagrams,mehta:decsion_tree_representation_of_boolean_functions,odonnell:decision_trees_influential_variable}. Of note is the significant savings possible by using semantic logic communication techniques; in the compression field, significant advances are often measured in few percentage point improvements whereas in here we are illustrating gains in integer multiples. On the right, we illustrate the setting where Bob does have access to a nontrivial logic fact $\receiverrndm$ that Alice doesn't know and Bob wants to prove all that Alice can prove ($\senderrndm=\queryrndm$). In red, we consider the same classic compression technique used in the previous plot, compared to the ultimate information theoretic bound.  As before, significant communication gains are possible.

{\bf No Need to Know.} A striking consequence of Theorem \ref{thm:core} is that, in this model, the same communication cost limit applies regardless of whether or not Alice knows what Bob knows. This phenomenon is also present in lossless classic communication \cite{slepian_wolf:coding} but it is more rare in the case of lossy compression \cite{wyner_ziv:coding, zamir:rate_loss_wz}. Logic semantic communication is inherently lossy, in that the logic statement reproduced by Bob need not resemble at all Alice's.

\begin{figure}
	\centering
\includegraphics[width=\textwidth]{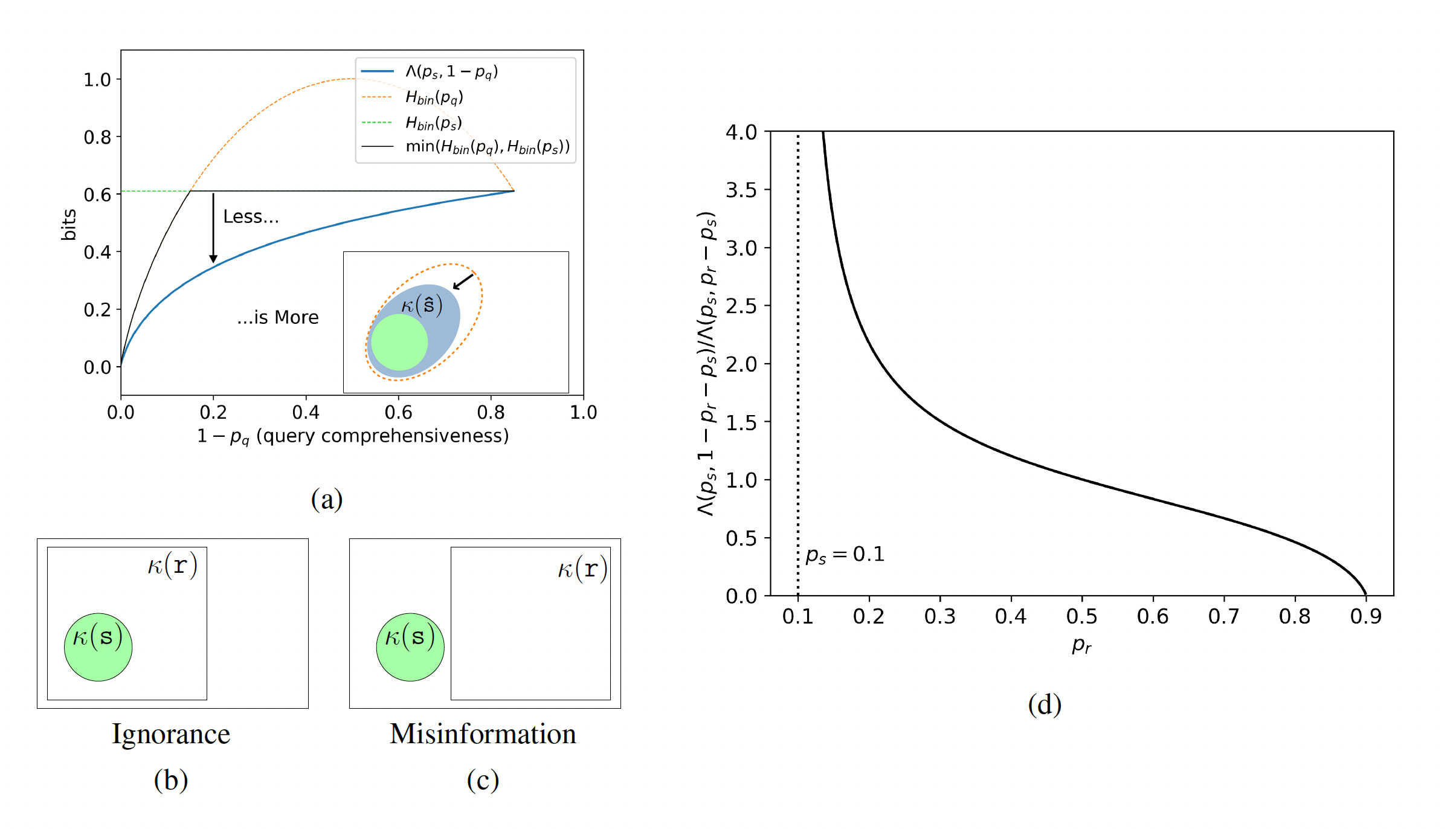} 
\caption{(\textbf{a}) \emph{Less is More.} In blue, the ultimate communication limit $\Lambda$  for the case $p_r=1$, as the query ranges from trivial ($p_q=1$) to coinciding with the sender's information ($p_q=p_s=0.15$). $\Lambda$ is cheaper (Less...) than the two obvious strategies, yet the kernel size received by Bob is smaller than that of the query, showing Bob can prove even more things (is More...) than required. A similar picture will hold for any $p_r$. (\textbf{b})-(\textbf{c}) \emph{Kernel diagrams} that model Bob's state of ignorance versus misinformation.
(\textbf{d}) \emph{Relative cost of misinformation to ignorance.} We compare the ultimate limits for (b) (ignorance) and (c) (misinformation); as Bob becomes more opinionated ($p_r \rightarrow p_s$, $p_s=0.1$), the ratio goes to infinity.}
\label{fig:no_need_less_more_misinf}
\end{figure}

{\bf The Less is More paradox.} This is the scenario of a query $\queryrndm$ that, while entailed by $\senderrndm$, is not necessarily equivalent to it ($p_q > p_s$). Our result holds in the case Alice knows $\receiverrndm$ for any $1 \geq p_r > p_q$.
In the
upper part
of Figure~\ref{fig:no_need_less_more_misinf}(a), we plot the information-theoretic limit in this scenario (in blue) and compare it against the bits that would be required to send either the query or the sender information, for the case $p_r=1$. The fundamental limit is lower than both of these (Less...). Yet, when one examines the nature of the solution, one realizes that Bob obtains a statement $\senderrndhat$ whose kernel is generally smaller than that of $\queryrndm$ demonstrating that Bob is able to prove more than what he needed (...is More). The paradox is explained by noting that the way this is achieved is through deciding, ahead of time, on a terse collection of approximating kernels each of which can be used to solve the problem for multiple combinations of $\senderrndm$ and $\queryrndm$, reducing further the total communication cost; a small concrete example showing this phenomenon is given in \cite{OurArxiv}. Of note, this also has potential implications for security -- being as efficient as one can to allow Bob to prove $\queryrndm$ using facts consistent with Alice's $\senderrndm$ results in revealing more than $\queryrndm$. In another play on words, one may say that one needs to say more to say less.

{\bf The price of misinformation.} In a line of work not directly covered by Theorem 1, but exploiting the same ideas, a rudimentary model of \emph{misinformation in a cooperative setting} is offered in \cite{OurArxiv}, defined as a setting where the logic statements of Alice and Bob are logically inconsistent (their kernels are non-overlapping), and additionally there's an agreement that Alice possesses the facts with which Bob wants to synchronize; we also make the simplifying assumption that Alice knows what Bob believes (Figure~\ref{fig:no_need_less_more_misinf}(c)). In contrast, in the setting that we have studied elsewhere in this article (Figure~\ref{fig:no_need_less_more_misinf}(b)), Bob may be said to be in a state of ignorance. Clearly this model is an unrealistic cartoon of the complex problem of misinformation; we introduce it here merely to explore the type of conclusions one can derive from it, fully aware of its limitations; see \cite{basu:semantic_extended} for an early occurrence of this problem. In the case of misinformation, the ultimate limit is $\Lambda(p_s,1-p_r-p_s)$, whereas for ignorance, as previously discussed, it is $\Lambda(p_s,p_r-p_s)$. We plot their ratio as a function of $p_r$ (Figure~\ref{fig:no_need_less_more_misinf}(d)) showing that as Bob becomes more opinionated ($p_r \rightarrow p_s)$, the relative cost of correcting misinformation vs.\ ignorance grows to infinity, a result that appears to agree with common intuition.

\subsection*{Implications and next steps}

We have provided evidence that new communication system designs incorporating deductive inference mechanisms can be effectively
analyzed and that the potential efficiency gains can be significant. To capitalize on this potential, nonetheless, additional progress is required. 

First, our deductive inference centered proposal can, in principle, be combined with other proposals \cite{liu_shao_zhang_poor:indirect_rd_semantic, liu:rd_semantic, guo2022semantic, stavrou:goal_semantic_rd} which similarly rely on Shannon-theoretic principles, potentially unveiling even more communication efficiencies.  Second, the communication model we introduced can be enhanced in many directions, including situations where Alice and Bob's roles are more symmetric (they each know something useful to a joint goal \cite{orlitsky:interactive}), adversarial (one may be trying to convince the other about something that is not true), consultant (Bob is seeking help for a specific matter), multiparty (one instructor, many students), and machine-oriented (Alice and Bob are AI agents). Practical algorithms for these scenarios, including better ones for those we already identified, can be obtained using additional ideas from coding theory (e.g., \cite{blahut1983algebraic,Richardson_Urbanke_2008})

We close with more speculative directions.  There is a significant opportunity for a semantic information theory to provide the same kinds of foundations for machine learning that standard information theory has for semantic AI models \cite{mackay2003}.  
In our full paper \cite{OurArxiv} we show that our results hold for Propositional Logic and First-Order Logic over finite models.
We anticipate that our results will be extended beyond finite model theory, and to richer logics beyond First-Order Logic, for example to First-Order Logic with counting \cite{IMMERMAN:1999}, Second-Order Logic \cite{fagin04, sep-logic-higher-order}, or to logic capturing uncertainty \cite{nilsson86, cozman00, fagin24}.
Given the close connections between Kolmogorov complexity and Shannon's entropy \cite{grunwald2004shannoninformationkolmogorovcomplexity}, we anticipate connections between semantic information theory and computer programs; we speculate on some of these in our longer paper. 

In the long term, our work, its extensions and particularly its combinations with other relevant disciplines may provide insights for the critical societal problems of efficient instruction and re-skilling, and countering misinformation and disinformation, for which society is currently struggling for solutions.



\clearpage 

%
\bibliography{scibib}

\bibliographystyle{sciencemag}


%
%
%
%
%
%


\section*{Acknowledgments}
The authors acknowledge helpful conversations with the following individuals: Ron Fagin, Phokion Kolaitis, Jason Rute, Kush Varshney and Mark Wegman.

\paragraph*{Funding:}
 Work of W. Szpankowski was partially supported by the NSF Center for Science of Information (CSoI) Grant CCF-0939370, and also by NSF Grants CCF-2006440, and CCF-2211423.




\newpage


\renewcommand{\thefigure}{S\arabic{figure}}
\renewcommand{\thetable}{S\arabic{table}}
\renewcommand{\theequation}{S\arabic{equation}}
\renewcommand{\thepage}{S\arabic{page}}
\setcounter{figure}{0}
\setcounter{table}{0}
\setcounter{equation}{0}
\setcounter{page}{1} 



\end{document}